\documentclass[preprint,aps,amsmath,superscriptaddress,nofootinbib,tightenlines]{revtex4}
\usepackage{graphicx,psfrag}
\usepackage{bm}
\usepackage{epsfig}

\def\OMIT#1{}

\newcommand{\nn}{\nonumber}

\newcommand{\bea}{\begin{eqnarray}}
\newcommand{\eea}{\end{eqnarray}}

\newcommand{\bnk}{\bar n \!\cdot\! k}
\newcommand{\bnq}{\bar n \!\cdot\! q}

\newcommand{\bnP}{\bar {\cal P}}

\newcommand{\nq}{n \!\cdot\! q}
\newcommand{\nk}{n \!\cdot\! k}

\newcommand{\jpsi}{J/\psi}

\begin{document}

%



\title{SCET approach to top quark decay}
	
\author{Xiaohui Liu}
\affiliation{Department of Physics and Astronomy,
	University of Pittsburgh,
        Pittsburgh, PA 15260\vspace{0.2cm}}

\date{\today\\ \vspace{1cm} }



\begin{abstract}
In this work we study QCD corrections to the top quark doubly decay rate with a detected
$B$ hadron containing a $b$ quark. We focus on the regime among which the emitted $W$ boson nearly carries its maxim energy. The tool that we use here is the soft-collinear effective theory (SCET). The factorization theorem based on SCET indicates a novel fragmenting jet function. We calculate this function to next-to-leading order in $\alpha_s$. Large logarithms due to several well separated scale are summed up using the renormalization group equation (RGE). Finally we reach an analytic formula for the distribution which could easily be generalized to other heavy hadron decay.
\end{abstract}

\maketitle

\newpage

\section{Introduction}
Top quark physics is one of the main subjects in theoretical and experimental particle physics~\cite{Beneke:2000xiv}. Recently an interesting proposal~\cite{Kharchilava:2000plb} has been suggested that top quark mass can be accurately measured
by studying top quark decays to to an exclusive hadronic state, for example $t\to W^++ B(b) \to W^+ + \jpsi$. For the sake of performing accurate studies of the top quark properties, a reliable description
of the distribution for top quark decay accompanied with bottom quark fragmentation is required. Unlike inclusive quantities, for analyses that require a detailed description of final states large logarithmic contributions arise due to the fact that the cancellation between infrared and ultraviolet divergence is not clean. These large logarithms must be resummed to all orders to make sensible predictions.
For processes with highly energetic hadron jets involved, a theoretical framework called soft collinear effective theory (SCET)~\cite{Bauer:2000prd, Bauer:2001prd, Bauer:2001plb, Bauer:2002prd} has the ability 
to sum up all those large logarithmic enhanced corrections. 

In our case, we consider the doubly decay rate ${\mathrm{d}^2\Gamma}/{\mathrm{d}y\mathrm{d}z}$, with $z$ is the energy fraction carried
by the $B$ hadron in the rest frame of the top quark and $y = m_{XB}^2/(m_t^2-m_W^2)$ being proportional to the invariant mass of the jet including the $B$ hadron. $y\to 0$ and $z \to 1$ correspond to
collinear and soft limit, respectively. We focus on the region which
$y\to 0$ but $z$ is around its intermediate region (neither close to $1$ nor to $0$). In this situation, the hadronic jet including the B meson is 
highly energetic and can be treated as massless. At this limit, $m^2_{XB}= 2q_B \cdot k_X$, thus $y$ can be related to the HERWIG~\cite{HERWIG}  variable $\xi$ by $y = (1+r)^2/2z(1-z)\xi$, where $r$ is the ratio of the $W$ boson mass to the top quark mass.  In SCET, a factorization theorem can be derived in a similar manner as the $B\to K X\gamma$ case~\cite{Procura:2010prd}
\bea\label{FACT}
\frac{\mathrm{d}^2\Gamma}{\mathrm{d}y\mathrm{d}z} 
= \Gamma_0 |C_H|^2 \frac{m_t^2(1-r)^2}{16\pi^3} \int_0^{p^+_{XB}}
\mathrm{d}k^+ {\cal G}_b^B(m_t(1-r^2)k^+,z,\mu)S_t(p_{XB}^+-k^+,\mu)\,,
\eea
where $r = m_W/m_t$, $p_{XB}^+ = m_t(1-r)/(1+r)y$ and $S_t$ is a soft function to describe the soft
nonperturbative gluons emitted by the top quark. $\Gamma_0$ is the decay rate at tree level which is 
\bea
\Gamma_0 = \frac{G_F m_t^3}{8\sqrt{2}\pi}(1-r^2)^2(1+2r^2) \,.
\eea
One interesting piece in the factorization theorem Eq.~(\ref{FACT}) is the fragmenting jet function~\cite{Procura:2010prd}, which naturally arises under SCET scheme. Compare to the traditional fragmentation function, the fragmenting jet function incorporate additional information about the invariant mass of the jet. Performing 
an operator product expansion, the fragmenting jet function can be
written as a convolution of a perturbatively calculable coefficient
${\cal T}$ and the standard fragmentation function. Ignoring mixing, this gives
\bea
{\cal G}^B_{b} (t,z,\mu) = \int_z^1 \frac{\mathrm{d}x}{x}\,
{\cal T}_{bb}\left(t,\frac{z}{x},\mu \right) D_b^B(x,\mu) \,.
\eea
And we note that the fragmentation function $D(z)$ can be further
 factorized into a convolution of a perturbative coefficient and a non-perturbative function. 

In Section~\ref{M}, we determine the coefficient $C_H$ and ${\cal T}_{bb}(t,z)$ by matching between different effective theories. In Section~\ref{R}, we use the REG to sum up large logarithmic contributions to derive an analytic formula for the doubly decay distribution.

\section{Matching}\label{M}
\begin{figure}[t]
\begin{center}
\includegraphics[width=5cm]{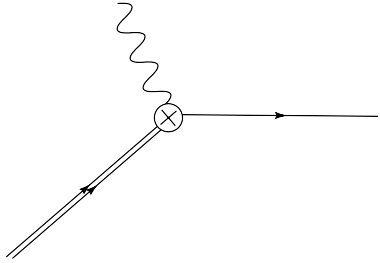}
\end{center}
\caption{\small
Tree level Feynman diagram for $t\to b+W^+$ in both QCD and SCET. 
Here the double line is an 
incoming top quark, single line stands for the b quark and the $W$ boson is given by the wavy line. 
\label{tdecay}}
\end{figure}

In this part, we calculate the coefficients $C_H$ and ${\cal T}_{bb}$ in Eq.~(\ref{FACT}) via matching. The leading order in power counting SCET operators contribute
to the process shown in fig.~\ref{tdecay} is given by
\bea
\sum_{i=0}^{2}\sum_{\omega} C_i(\omega) \bar{\xi}_n W_n \delta_{\omega,\bnP^\dagger} \,
\Gamma_i^\mu \,
Y^\dagger h_v \,.  
\eea
where $\xi_n$ is the collinear light quark propagating in the light cone direction $n$ and $h_v$ is the field annihilating a heavy quark with velocity $v$. $W_n$ is the collinear Wilson line built out of 
collinear gauge field, which is essential in constructing gauge invariant operators in SCET~\cite{Bauer:2001plb} and $Y$ is the usoft
Wilson line emerges from decoupling the usoft gluons from the leading order collinear modes~\cite{Bauer:2002prd}, which is crucial in deriving the factorization therom Eq.~(\ref{FACT}). $\bnP$ is an operator which picks out large label momentum~\cite{Bauer:2001plb}.

The basis for the Dirac structures are
\bea
\Gamma_0^\mu = \gamma^\mu P_L\,,\hspace{3.ex} \,
\Gamma_1^\mu = \frac{n^\mu}{n\cdot v} P_R \,,  \hspace{3.ex} \,
\Gamma_2^\mu = v^\mu P_R \,,
\eea
where $P_{R/L} = (1 \pm \gamma^5)/2$. $C_1$ and $C_2$ vanish at tree level.

Now we match QCD amplitude onto the ${\rm SCET}_{\rm I}$ operators to one loop. We calculate the virtual corrections to the 
${\rm SCET}_{\rm I}$ current at
the order $\alpha_s$, then comparing with the QCD amplitude at 
the same order~\cite{Campbell:2004prd}, we determine the Wilson coefficients and the 
matching scale $\mu_H$, as well. We should expect 
that the ${\rm SCET}_{\rm I}$ calculations 
reproduce the infrared divergence in QCD.

\begin{figure}[t]
\begin{center}
\includegraphics[width=9cm]{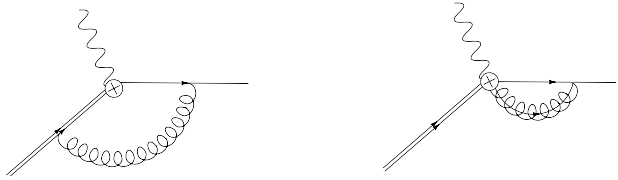}
\end{center}
\caption{\small
QCD virtual corrections to the ${\rm SCET}_{\rm I}$ operator at the order 
${\cal O}(\alpha_s)$. The spring line is a usoft gluon and
the collinear gluon are represented by a spring with a line
going through. 
\label{vc}}
\end{figure}

The leading order QCD virtual corrections to the ${\rm SCET}_{\rm I}$ operator 
are shown in fig.~\ref{vc} except for the self energy corrections. 
Once we ignore the $b$ quark mass,
the loop integrals are scaleless and vanish in dimensional regularization. In order to extract ultraviolet divergence, we
put $b$ quark offshell here. Evaluating those diagrams in 
$d =4-2\epsilon$ dimensions gives divergences from usoft 
vertex correction
\bea
I_{\rm usoft}= -\frac{\alpha_sC_F}{4\pi}\,
\left(\frac{1}{\epsilon^2} - \,
\frac{2}{\epsilon}\log\left(-\frac{\nq}{\mu}\right) \right)\,
{\cal O}_0\,,
\eea
as well as the collinear gluon correction
\bea
I_{\rm coll}=-\frac{\alpha_sC_F}{4\pi}\,
\left(\,
-\frac{2}{\epsilon^2}-\frac{2}{\epsilon}+\frac{2}{\epsilon}\,
\log \left(-\frac{\bnq \nq}{\mu^2} \right) \,
\right)\,
{\cal O}_0\,.
\eea
Here, $q$ is the momentum carried by the outgoing $b$ quark.

The summation of the divergent piece should be canceled by the
operator counterterm $\delta Z_{\cal O}$ 
together with the wavefunction counterterms. Since 
$\delta Z_t = \alpha_s C_F/(2\pi \epsilon)$ and $\delta Z_b = -\alpha_s C_F/(4\pi\epsilon)$ for heavy and collinear quark wavefunction conterterms, respectively, we can extract $\delta Z_{\cal O}$,
\bea
\delta Z_{\cal O} = \frac{\alpha_sC_F}{4\pi}\,
\left( \frac{1}{\epsilon^2} \,
+\frac{5}{2\epsilon}
-\frac{2}{\epsilon}\log \left(\frac{\bnq}{\mu} \right)\,
\right)\,.
\eea
Thus, in ${\rm SCET}_{\rm I}$ the leading order plus 
one-loop virtual correction to the differential decay rates is
\bea
\frac{\mathrm{d}^2\Gamma^{\rm I}_{{\rm L}+{\rm V}}}{\mathrm{d}y \mathrm{d}z}\,
= \Gamma_0 \left(1+C_0+\frac{C_1}{2}\frac{1-r^2}{1+2r^2}\right)\,
 \delta(y)\delta(1-z) \,
\left(1-\frac{\alpha_sC_F}{2\pi}\,
\left( \frac{1}{\epsilon^2} \,
+\frac{5}{2\epsilon}
-\frac{2}{\epsilon}\log \left(\frac{\bnq}{\mu} \right)\,
\right) \right) \,.
\eea
We see that the ${\rm SCET}_{\rm I}$ result reproduces
exactly the same infrared poles in QCD~\cite{Campbell:2004prd} 
as expected and
the matching coefficent $C_0$ and $C_1$ are
\bea\label{mc}
&&C_0 =  \frac{\alpha_sC_F}{2\pi}  \,
\left( -\frac{1}{2} \log\frac{\mu^2}{m_t^2}  \,
\left(\log\frac{\mu^2}{m_t^2(1-r^2)^4} +5 \right)
-\frac{\pi^2}{4} - 6 \right. \nn \\
&&\left. \hspace{10.ex} -2{\rm Li}_2(r^2) -2\log^2(1-r)^2 - \frac{1-3r^2}{r^2}\log(1-r^2)  \,
\right)\,, \nn \\
&&C_1 =  \frac{\alpha_sC_F}{2\pi} \frac{2}{r^2} \log(1-r^2) \,.
\eea
We choose the hard matching scale be $\mu_H = \bnq = m_t(1-r^2)$ to eliminate large logarithms.

Now we turn to the matching between ${\rm SCET}_{\rm I}$ and
${\rm SCET}_{\rm II}$, which will determine the coefficient 
${\cal T}(t,z)$ in the fragmenting jet function. The matching is
done at decay rate level at the limit $y\to 0$. Thus the coefficient
is dominated by those singular terms in this limit.

\begin{figure}[t]
\begin{center}
\includegraphics[width=9cm]{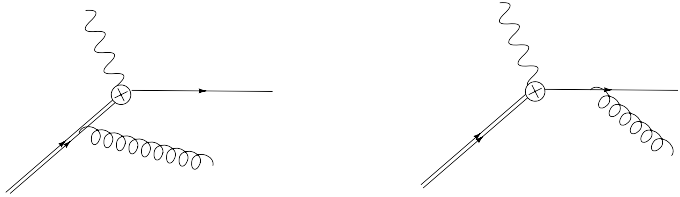}
\end{center}
\caption{\small
Real emission of a usoft gluon in {\rm SCET}.
\label{semit}}
\end{figure}

\begin{figure}[t]
\begin{center}
\includegraphics[width=9cm]{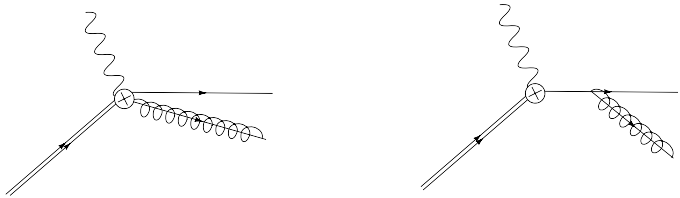}
\end{center}
\caption{\small
Real emission of a collinear gluon in {\rm SCET}.
\label{cemit}}
\end{figure}

The diagrams for usoft and collinear real emissions at next-to-leading-order in $\alpha_s$ are shown in fig.~\ref{semit} and fig.~\ref{cemit}, respectively. The amplitude square 
coming from the usoft emission is the same as making the eikonal approximation in QCD which gives
\bea
|{\cal M}|_{\rm usoft}^2 = g_s^2 C_F  |{\cal M}|_0^2 \,
\left(\frac{2}{\nk v\cdot k} - \frac{1}{(v\cdot k)^2} \right) \,,
\eea
where $k$ is the momentum for the real gluon emitted. 

The collinear diagrams can be evaluated using the SCET 
Feynman rules~\cite{Bauer:2001prd}.
However at certain regions of the phase space, for
example when $y \to 0$ while $z \to 1$, 
 the collinear gluon momentum $k$ will become usoft and scales like 
$Q(\lambda^2,\lambda^2,\lambda^2$) rather 
than $Q(\lambda^2,1,\lambda)$.
In this regime, the SCET diagrams will include a double power counting. To get rid of double counting, we should subtract the 
"zero-bin" contribution~\cite{Manohar:2007prd} from the collinear diagrams. 
In our case, the zero-bin can be calculated simply by treating 
the gluon with momentum $k$ in fig.~\ref{cemit} as a usoft mode.   
After perform the zero-bin subtraction, collinear real 
emission is given by
\bea
|{\cal M}|_{\rm coll}^2 = \,
g_s^2C_F |{\cal M}|^2_0 \frac{1}{q_{gb}^2}\left( \,
\frac{4\bnq}{\bnk}+(2-2\epsilon)\frac{\nq}{{\nq}_{gb}} \,
- \frac{4{\bnq}_{gb}}{\bnk}\,
\right) \,,
\eea
where $q$ is the $b$ quark momentum and $q_{gb}$ is the total
momentum for the $b$ quark-gluon system. The last term in the equation
above corresponds to the zero-bin subtraction.


Combining usfot, collinear and zero subtraction, we calculate the differential decay rates in ${\rm SCET}_{\rm I}$, which yields
\bea\label{dSCETI}
\frac{\mathrm{d}^2\Gamma^{\rm I}_{\rm R}}{\mathrm{d}y \mathrm{d}z}\,
& = &\frac{\alpha_s C_F}{2\pi} \Gamma_0 |C_{\rm H}|^2 \,
\left( \frac{4\pi\mu^2}{m_t^2(1-r^2)^2} \right)^\epsilon\,
\frac{(1+r)^{2\epsilon}}{\Gamma(1-\epsilon)}\,
z^{-\epsilon}(1-z)^{-\epsilon} \,
\left(\frac{y}{y_{max}}\right)^{-\epsilon}\,
(y_{max} - y)^{-\epsilon} \nn \\
&& \times \left[\frac{1}{y}\,
\left(\frac{z^2+1 -\epsilon(1-z)^2}{1-z}\right)\,
-\frac{2}{(1+r)^2}\frac{1}{(1-z)^2} \right]\,,
\eea
where all the hard matching coefficients in Eq.~(\ref{mc}) are included
in $|C_{\rm H}|^2$. 

To determine the coefficient ${\cal T}(t,z)$ in the fragmentng
jet function, we compare the cross section calculated within 
${\rm SCET}_{\rm I}$ and the one in ${\rm SCET}_{\rm II}$. The matching
procedure is similar to Ref.~\cite{Fleming:2006prd}
. However, in our case, extracting the 
singular contributions is 
complicated due to the fact that $y_{max}$ is not linear in $z$. A simple way to do the matching is based on the fact that the fragmenting jet function is universal and in principle itself has
 no information about the $W$ boson mass, thus, 
formally this function doesn't 
depend on $r$ explicitly. This allows us to set $r$ to $0$ to simplify 
the calculation. (In this case, $y_{max} = 1-z$ which is identical to Ref.~\cite{Fleming:2006prd}) After obtaining the coefficient ${\cal T}(t,z)$, we then restore the $r$ dependence. 
 
Here, we keep the $r$ dependence explicitly. We slightly generalize
the method proposed in Ref.~\cite{Fleming:2006prd} to investigate
the singular behavior as $y \to 0$ in Eq.~(\ref{dSCETI}) in the 
Appendix. The virtual corrections to the cross section should also
be included to this order. Since the loops are scaleless and thus 
vanish in dimensional regularization. Therefore the infrared divergent part is the same as minus the counterterm. Once including both real
and virtual corrections, we find that in ${\rm SCET}_{\rm I}$
\bea\label{dSCETIo}
\frac{\mathrm{d}^2\Gamma^{\rm I}_{{\rm R}+{\rm V}}}{\mathrm{d}y \mathrm{d}z}\,
& = &\frac{\alpha_s C_F}{2\pi} \Gamma_0 |C_{\rm H}|^2 \,
 \left\{ \,
\delta(y)\delta(1-z)
\left(\,
-\log\left(\frac{\mu_H^2}{\mu^2}\right)
+\frac{1}{2}\log^2\left(\frac{\mu_H^2}{\mu^2}\right)\,
 -\frac{\pi^2}{4}\right)\,
  \right. \nn \\
&& \left. + \delta(y)\left[- \frac{1}{\epsilon}P_{qq}(z)\,
+ \bar{P}_{qq}(z)\log\left(z \right) \,
+ (1+z^2)\left(\frac{\log(1-z)}{1-z}\right)_+ + (1-z) \right] \right.\nn \\
&& \left. -2\left[ \kappa\left( \frac{1}{\kappa y} \right)_+\,
+ \kappa\left(\frac{\log(\kappa y)}{\kappa y} \right)_+\,
\right] \delta(1-z)+ \frac{\kappa\mu_H^2}{\mu^2} \left(\frac{1}{\frac{\kappa\mu_H^2}{\mu^2}y}  \right)_+ {\bar P}_{qq}(z)  \, 
\right\} \,,
\eea
where, we define $\kappa = 1/(1+r)^2$ and $\mu_H = m_t(1-r^2)$. 
We have used the identity Eq.~(\ref{idd}) for the plus-prescription in the Appendix. Here
\bea
P_{qq}(z) = \frac{1+z^2}{(1-z)_+} + \frac{3}{2}\delta(1-z)
= \bar{P}_{qq}(z) + \frac{3}{2} \delta(1-z) \,,
\eea
is the quark to quark splitting function.

In ${\rm SCET}_{\rm II}$ the decay rates read as
\bea\label{dSCETII}
\frac{\mathrm{d}^2\Gamma_{\rm II}}{\mathrm{d}y \mathrm{d}z}\,
= \Gamma_0 |C_{\rm H}|^2 \frac{m_t^2(1-r)^2}{2(2\pi)^3}\,
\int_0^{p_{gb}^+} \mathrm{d}k^+ \int_z^1 \frac{\mathrm{d}x}{x} \,
{\cal T}_{bb}\left(\mu_H k^+,\frac{z}{x}\right)D_b(x) S_t(p^+_{gb}-k^+) \,.
\eea
By definition, $p_{gb}^+ = m_t(1-r)/(1+r)y$ and we suprress the scale
dependence here. We can perform expansions for those functions involved in the differential decay rates to order $\alpha_s$,
\bea
&&{\cal T}_{bb}(\omega k^+,z)  = 2(2\pi)^3\left(\,
\delta(\omega k^+) \delta(1-z)\,
+ \frac{\alpha_sC_F}{2\pi}{\cal T}_{bb}^{(1)}(\omega k^+,z)\right) \,, \nn \\
&&S_t(k^+) = \delta(k^+) + \frac{\alpha_sC_F}{\pi}S^{(1)}_t(k^+)\,,\nn\\
&&D_b(x) = \delta(1-x) - \frac{\alpha_sC_F}{2\pi \epsilon}P_{qq}(x) \,.
\eea
Therefore, omitting the leading term in $\alpha_s$, 
we can manipulate Eq.~(\ref{dSCETII}) to the form
\bea\label{dgSCETII}
\frac{\mathrm{d}^2\Gamma^{(1)}_{\rm II}}{\mathrm{d}y \mathrm{d}z}\,
= \,
\frac{\alpha_sC_F}{2 \pi }\Gamma_0 |C_{\rm H}|^2   \left[ \,
\delta(y)\frac{-P_{qq}(z)}{\epsilon} \,
+ 2\delta(1-z)(\kappa \mu_H)  S_t^{(1)}(\kappa \mu_H y )\,
+ (\kappa \mu^2_H) {\cal T}_{bb}^{(1)}(\kappa \mu_H^2 y,z)\right]  \,.
\eea

The shape function here is the same as the one in $B$ meson decay which
has been calculated in Ref.~\cite{Bauer:2000prd}. We follow their procedure to get
\bea\label{shape}
(\kappa \mu_H)S_t^{(1)}(\kappa \mu_H y) &=& 
\delta(y)\left( -\frac{1}{2}\log\left(\frac{\mu_H^2}{\mu^2}\right)\,
+\frac{1}{4}\log^2\left(\frac{\mu_H^2}{\mu^2}\right)\,
-\frac{\pi^2}{24}\,
\right)\,  \nn \\
&& \hspace{2.ex} -\kappa\left( \frac{1}{\kappa y} \right)_+ \,
-\kappa\left( \frac{\log(\kappa y)}{\kappa y} \right)_+ \,
-\frac{\kappa \mu_H^2}{\mu^2}\,
\left( \frac{\log(\frac{\kappa \mu_H^2}{\mu^2}y)}{\frac{\kappa \mu_H^2}{\mu^2}y} \right)_+ \,.
\eea
Plugging Eq.~(\ref{shape}) into Eq.~(\ref{dgSCETII}) and comparing
with the decay rate in ${\rm SCET}_{\rm I}$ Eq.~(\ref{dSCETIo}), we
can derive the coefficient ${\cal T}_{bb}$
\bea\label{tj}
&&{\cal T}^{(1)}_{bb}(t,z) = \,
\delta(t)\left(\,
\bar{P}_{qq}(z)\log(z) +(1+z^2)\left( \frac{\log(1-z)}{1-z} \right)_+ + (1-z) 
 -\frac{\pi^2}{6}\delta(1-z) \right) \nn \\
&& \hspace{15.ex}  +\frac{1}{\mu^2}\left(\frac{1}{t/\mu^2} \right)_+\bar{P}_{qq}(z) + \,
 \frac{2}{\mu^2}\left(\frac{\log(t/\mu^2)}{t/\mu^2} \right)_+ \delta(1-z)\,.
\eea
Here $t = \kappa \mu_H^2 y$ is the invariant jet mass. Requiring all
large logarithms to vanish, the intermediate matching scale should
be set to the jet mass, $\mu_c^2 = t$. And we see from Eq.~(\ref{tj}) that formally the matching coefficient can not depend on $r$ as we 
explained before. We can check that after integrating Eq.~(\ref{tj}) over $z$, we recover the massless collinear quark jet function at order $\alpha_s$ in SCET.

\section{Running}\label{R}
The differential decay rate has several well separated scales 
$\mu_H$, $\mu_c$ and $\mu_s$ involved. To go from one scale to another,
we use the renormalization group equation to sum up large logarithms.
First the ${\rm SCET}_{\rm I}$ operators are run from hard scale $\mu_H, $ using the ${\rm SCET}_{\rm I}$ RGEs, down to the collinear
scale $\mu_c$ at which ${\rm SCET}_{\rm I}$ is matched onto ${\rm SCET}_{\rm II}$. Then we run the shape function to the scale 
$\mu_s = \mu_c^2/\mu_H$. 

There are several ways to perform this procedure~\cite{Leibovich:2000prd, Becher:2006prl, Fleming:2008prd}. We choose to do the running in the moment space then by take the inverse Mellin transform to obtain a resummed decay rate~\cite{Leibovich:2000prd}. 
In the moment space the formula for the decay rate could be written as
\bea
\Gamma_N = \Gamma_0|C_H(\mu_c)|^2 
\int_z^1 \frac{\mathrm{d}x}{x}\,
\hat{{\cal T}}\left(\frac{z}{x},N,\mu_c\right)D_b(x,\mu_c) \,
\hat{S}_t(N,\mu_s) \,.
\eea
To obtain the moment space decay rate above, we first normalize the 
fragmenting jet function and the shape function in a way that both
functions are dimensionless quantities, which we use hats to 
represent for. We define a variable $\bar{y} = 1- y$ and the moments are taken respect to $\bar{y}$. Also we introduce $u$ to express $k^+$ in Eq.~(\ref{dSCETII}) in terms of $\kappa \mu_H(1-u)$. In the regime $y\to 0,{\bar y}\to 1$, large $N$ limit is achieved. In moment space, the scales 
are $\mu_c = \mu_H\sqrt{\kappa}/\sqrt{\bar{N}}$ and $\mu_s = \kappa\mu_H/\bar{N}$. The hard scale
$\mu_H$ is the same as defined in the previous section. 

At the collinear scale $\mu_c$, the large logarithms in the
matching coefficient $\hat{{\cal T}}$ vanish, which gives
\bea\label{Ts}
\hat{{\cal T}}(z,N,\mu_c) = \delta(1-z)+\frac{\alpha_sC_F}{2\pi}\,
\left( \log(z) {\bar P}_{qq}(z) \,
 + (1+z^2)\left( \frac{\log(1-z)}{1-z}\right)_+   \,
+ (1-z)  \,
\right) \,.
\eea
The only $N$ dependence are through $\mu_c$ in the 
strong coupling $\alpha_s$.

Now we take another Merlin transform respect to $z$, 
\bea
\Gamma_{NM} = \Gamma_0 |C_H(\mu_c)|^2 \,
\left(1 + \frac{\alpha_s(\mu_c)C_F}{2\pi}T(M) \right) D_b(M,\mu_c)
\hat{S}_t(N,\mu_s) \,.
\eea
The running of the fragmentation function $D(M,\mu)$ in the moment
space is given by
\bea
\mu \frac{\mathrm{d}}{\mathrm{d}\mu} D(M,\mu) = \,
\frac{\alpha_s}{4\pi} a(M)D(M,\mu) \,.
\eea
The leading order solution is then
\bea\label{runD}
D(M,\mu_c) = D(M,\mu_H)  \,
\exp\left(\frac{a(M)}{2\beta_0}\log(1-\chi) \right) \equiv \,
D(M,\mu_H)\exp\left( h_M(\chi) \right) \,,
\eea
with $\chi=\log(\bar{N}/\kappa)\alpha_s(\mu_H)\beta_0/4\pi$ and $\beta_0 = (11C_A -2 n_f)/3$. To the leading order, the running of the combination 
$\alpha_sC_F/(2\pi)T(M)D(M,\mu)$ satisfies similar equation as Eq.~(\ref{runD}) with $a(M)$ replaced by $4a(M)-2\beta_0$. Therefore, we
can define $h'_M(\chi)$ in the same way as $h_M(\chi)$ and have
\bea
\frac{\alpha_s(\mu_c)C_F}{2\pi} T(M)D(M,\mu_c) = \frac{\alpha_s(\mu_H)C_F}{2\pi}T(M) D(M,\mu_H) \exp(h'_M(\chi)) \,.
\eea
All $M$ dependence has been moved into factor $h_M$ and $h'_M$. 

The running of the ${\rm SCET}_{\rm I}$ currents along with the 
shape function could be lifted from Ref.~\cite{Fleming:2006prd}. We obtain the following resummed decay rate in the moment space:
\bea
\Gamma_{NM} &= &\Gamma_0 |C_H(\mu_H)|^2 e^{\log(N/\kappa)g_1(\chi)+g_2(\chi)}\,
\hat{S}_t(N,\mu_s) \nn \\
&& \hspace{12.ex} \times \left(e^{h_M(\chi)}+e^{h'_M(\chi)}\frac{\alpha_s(\mu_H)C_F}{2\pi} T(M) \right)D_b(M,\mu_H) \,,
\eea
where
\bea
&&g_1(\chi) = -\frac{2\Gamma_1}{\beta_0\chi}\left[\,
(1-2\chi)\log(1-2\chi)-2(1-\chi)\log(1-\chi) \,
\right]\,,  \nn \\
&&g_2(\chi) = -\frac{8\Gamma_2}{\beta_0^2} \left[\,
-\log(1-2\chi)+2\log(1-\chi) \right] \nn \\
&&\hspace{8.ex} \,
-\frac{2\Gamma_1 \beta_1}{\beta_0^3} \left[ \,
\log(1-2\chi) - 2\log(1-\chi) + \frac{1}{2}\log^2(1-2\chi)\,
-\log^2(1-\chi)
\right] \nn \\
&&\hspace{8.ex}\,
+\frac{4\gamma_1}{\beta_0} \log(1-\chi) + \frac{2B_1}{\beta_0}\log(1-2\chi) \nn \\
&&\hspace{8.ex}\,
-\frac{4\Gamma_1}{\beta_0}\log n_0 \left[ \log(1-2\chi) - \log(1-\chi) \right]  \,,
\eea
with $n_0 = e^{\gamma_E}$ and 
\bea
&&\Gamma_1 = 4C_F \,, \hspace{5.ex} \Gamma_2 = C_A\left[C_A\left(\frac{67}{36}-\frac{\pi^2}{12}\right)-\frac{5n_f}{18}\right] \,, \nn \\
&&B_1 = -4 C_F \,, \hspace{5.ex} 2 \gamma_1 = -\frac{3}{2}C_F \,, \nn \\
&&\beta_1=\left(\frac{34}{3}C_A^2 - \frac{10}{3}C_An_f - 2C_F n_f\right) \,.
\eea

Evaluating the inverse Mellin transform with respect to $N$ using the 
results of Ref.~\cite{Leibovich:2000prd} shows that
\bea
\frac{\mathrm{d}\Gamma_M}{\mathrm{d}y} &=& \,
\Gamma_0|C_H(\mu_H)|^2 \int_{1-y}^1 \frac{\mathrm{d}u}{u}
\hat{S}_t\left(\frac{1-y}{u}\right)  \left[  \,
-u\frac{\mathrm{d}}{\mathrm{d}u} \,
\left(\theta(1-u) \frac{e^{lg_1(l)+g_2(l)}}{\Gamma\left[\,
1-g_1(l) - lg_1'(l) \right]}\, \right. \right. \nn \\
&& \hspace{9.ex} \times \left. \left.
\left(e^{h_M(l)}+e^{h'_M(l)}\frac{\alpha_s(\mu_H)C_F}{2\pi} T(M) \right)D_b(M,\mu_H) \,
\right)\,
\right]\,,
\eea
where $l = -\alpha_s \beta_0/(4\pi)\log(1-u)$ and 
$g_1'(l) = \mathrm{d}g_1(l)/\mathrm{d}l$. The factor $h_M(l)$ can be
eliminated using Eq.~(\ref{runD})
\bea
e^{h_M(l)}D(M,\mu_H) = \exp\left[\frac{a(M)}{2\beta_0} \log(1-l)\right]D(M,\mu_H) = D(M, \kappa \mu_H \sqrt{1-u}) \,,
\eea
and the same thing holds for $h'_M(l)$. 

After eliminating both factors $h_M$
and $h'_M$, all the $M$ dependence is now entirely 
included in the moments of the fragmentation function, so 
the inverse Mellin transform with respect to $M$ is straightforward. 
Hence we derive the resummed decay rate:
\bea\label{decayfnl}
\frac{\mathrm{d}^2\Gamma}{\mathrm{d}y\mathrm{d}z} \,
&=& \Gamma_0|C_H(\mu_H)|^2 \,
\int_{1-y}^1 \frac{\mathrm{d}u}{u}
\hat{S}_t\left(\frac{1-y}{u}\right)  \left[  \,
-u\frac{\mathrm{d}}{\mathrm{d}u} \,
\left(\theta(1-u) \frac{e^{lg_1(l)+g_2(l)}}{\Gamma\left[\,
1-g_1(l) - lg_1'(l) \right]}\, \right. \right. \nn \\
&& \hspace{9.ex} \times \left. \left.
\left( \delta(1-z) + 
\frac{\alpha_sC_F}{2\pi}\tilde{T}^{(1)}(z)\right) \otimes D_b(z,\kappa \mu_H \sqrt{1-u}) \,
\right)\,
\right]\,,
\eea
where the convolution is defined as 
$f\otimes g = \int_z^1\mathrm{d}x/x f(x)g(x/z)$ and $\tilde{T}^{(1)}(z)$ is the second term in Eq.~(\ref{Ts}). We note that in the second
line the $\alpha_s$ has an scale dependence on $\kappa \mu_H \sqrt{1-u}$
which has been suppressed. Due to the universality of the fragmenting jet function, Eq.~(\ref{decayfnl}) can also be applied to other processes like heavy meson decay $B\to XK\gamma$ and etc. When applying Eq.~(\ref{decayfnl}), we should be careful in dealing with the Landau poles since the functions $g_i(l)$ blow up as $u$ approach $1$. A simple way to avoid
Landau pole is to set an upper limit on $u$. And it has been argued that the difference between integrating to this upper limit $u_{max}$ and to one is of order power suppressed corrections~\cite{Leibovich:2001plb}.

\section{Summary}
We have discussed the top quark doubly differential decay rate near
the phase space boundary where the $W$ boson carries its maxim energy within the framework of soft collinear effective theory. The factorization theorem for top quark decay is similar to the one
for $B\to XK\gamma$, in which a novel fragmenting jet function arises in replacement of the standard parton fragmentation function. The fragmenting jet function provides information on the invariant mass
of the jet from which a detected hadron framents. In this work we calculated the fragmenting jet function to next-to-leading order in $\alpha_s$ by comparing the decay rates calculated in ${\rm SCET}_{\rm I}$ and ${\rm SCET}_{\rm II}$. We also check the relation between our
derived fragmenting jet function with the inclusive collinear quark jet function, finding that they satisfy ${\cal J}(t) \to {\cal G}(t,z)\mathrm{d}z$ as indicated in Ref.~\cite{Procura:2010prd}. We use the
renormalization group equation to sum up large logarithms involved 
in the decay rates. After resummation, we arrive at an analytic formula for the distribution. Our results can be applied to other heavy hadron decay processes with a detected light hadron like $B$ meson radiative decay. And the result of this work may help tuning event generators such as Herwig.

\acknowledgements
We would like to thank A.~K.~Leibovich and A.~Freitas for discussions.
X.L. was supported by the National Science Foundation under Grant
No. Phy-0546143.

\section*{APPENDIX}

In this section, we show how to extract contributions which are singular as $y\to 0$. And for this reason we will drop all terms which are regular in this limit.

First we consider the combination of the form
\bea
I_1[y,z]\equiv (1+r)^{2\epsilon}\, y^{-1-\epsilon}\,
\frac{y_{max}^{\epsilon}(y_{max}-y)^{-\epsilon}}{(1-z)^{1+\epsilon}}\,. 
\eea
where $y_{max} = (1+r)^2z(1-z)/(z+r^2(1-z))$. 

We start with considering the integration
\bea\label{disapp}
\int_{z_{min}}^{z_{max}} \mathrm{d}z I_1[y,z]\left(g(z)-g(1)\right)\,
 + g(1)  \int_{z_{min}}^{z_{max}}\mathrm{d}z  I_1[y,z]\,, 
\eea
with $z_{max} = 1 - 1/(1+r)^2 y +{\cal O}(y^2) $ and $z_{min} = {\cal O}(y)$. So $z_{max}$ goes to $1$ as $y$ goes to $0$ while $z_{min}$ approaches to $0$ in this limit. 

Due to the distributional identity,
\bea
\frac{1}{y^{1+\epsilon}} = - \frac{1}{\epsilon}\delta(y)\,
+\left( \frac{1}{y} \right)_+ \,
-\epsilon \left( \frac{\log y}{y} \right)_+ \,,
\eea
the non-singular contributions as $y \to 0$, including the integration limits, in the first term of Eq.~(\ref{disapp}) could be expanded 
around $y=0$ and leaves out all terms of order ${\cal O}(y)$ or higher.
Thus the first term becomes
\bea
&&(1+r)^{2\epsilon} \, y^{-1-\epsilon}\,
\int_0^1 \mathrm{d}z \frac{g(z)-g(1)}{(1-z)^{1+\epsilon}} \, \nn \\
&&\hspace{6.ex} = (1+r)^{2\epsilon} \, y^{-1-\epsilon}\,
\int_0^1 \mathrm{d}z \left[\left(\frac{1}{1-z} \right)_+ \,
- \epsilon \left(\frac{\log(1-z)}{1-z}\right)_+ \right] g(z)\,.
\eea
Using the distributional identity and expand in $\epsilon$ gives
that 
\bea
&&\int_{z_{min}}^{z_{max}}\mathrm{d}z I_1[y,z]\left(g(z)-g(1)\right)\nn\\
&=&\int_0^1\mathrm{d}z\left\{ \delta(y)\left[-\frac{1}{\epsilon}\left(\frac{1}{1-z} \right)_+\,
+ \left( \frac{\log(1-z)}{1-z}\right)_+  \right] \,
+ \kappa \left( \frac{1}{\kappa y}\right)_+  \left( \frac{1}{1-z}\right)_+ \right\} g(z) \,.
\eea
Here $\kappa = 1/(1+r)^2$ and we have applied the relation
\bea\label{idd}
\kappa \left(\frac{\log^n(\kappa y)}{\kappa y} \right)_+ =\,
 \frac{\log^{n+1}(\kappa)}{n+1 } \delta(y) \,
+ \sum_{k=0}^n \frac{n!}{(n-k)!k!}\log^{n-k}(\kappa) \,
\left( \frac{\log^k(y)}{y}\right)_+
\eea
Now we turns to the second term in Eq.~(\ref{disapp}) by considering
a further integration 
\bea\label{distapp}
\int_0^1 \mathrm{d}y \int_{z_{min}}^{z_{max}} \mathrm{d}z \,
I_1[y,z]\left(f(y)-f(0) \right) + f(0) \int_0^1 \mathrm{d}y \int_{z_{min}}^{z_{max}} \mathrm{d}z I_1[y,z] \,.
\eea
Since the first term in the equation above is finite as $y \to 0$. 
We can set $\epsilon = 0$ and then perform the integration over $z$, 
which results in 
\bea
&&\int_0^1 \mathrm{d}y \int_{z_{min}}^{z_{max}} \mathrm{d}z \,
I_1[y,z]\left(f(y)-f(0) \right) \nn \\
&&\hspace{6.ex} = - \int_0^1 \mathrm{d}y \frac{1}{y}\,
\log\left( \frac{1-z_{max}}{1-z_{min}}\right)\,
\left(f(y)-f(0)\right) \nn \\
&&\hspace{6.ex} = - \int_0^1 \mathrm{d}y \,
\kappa \left(\frac{\log(\kappa y)}{\kappa y}\right)_+ f(y)\,.
\eea
The last equation is obtained by expanding $z_{max}$ and $z_{min}$
around $y=0$, e.g., 
$\log(1-z_{max}) = \log(\kappa y (1+{\cal O}(y))) = \log(\kappa y)+ {\cal O}(y)$, 
and ignore all the non-singular contributions in $y$.

Evaluating the integration of the second term in 
Eq.~(\ref{distapp}) gives 
\bea
f(0) \int_0^1 \mathrm{d}y \int_{z_{min}}^{z_{max}} \mathrm{d}z I_1[y,z] 
= \left(\frac{1}{2 \epsilon^2} -\frac{\pi^2}{12} \right)f(0) \,.
\eea

Gathering all the pieces, we have
\bea
&&\int_{z_{min}}^{z_{max}}\mathrm{d}z I_1[y,z] g(z) \nn\\
&=& \int_0^1 \mathrm{d}z \left[ \,
\delta(y)\left( \,
\left(\frac{1}{2 \epsilon^2} -\frac{\pi^2}{12}\right)\delta(1-z)\,
 -\frac{1}{\epsilon} \left(\frac{1}{1-z} \right)_+ \,
+\left( \frac{\log(1-z)}{1-z}\right)_+ \, \right) \right.\nn \\
&&\left.  + \kappa \left(\frac{1}{\kappa y} \right)_+\,
\left( \frac{1}{1-z}\right)_+  \,
- \kappa \left(\frac{\log(\kappa y)}{\kappa y} \right)_+ \delta(1-z)
 \right] g(z) \,.
\eea

Next we consider another integration which will contribute to the 
non-singular part as $y$ goes to $0$
\bea
\int_{z_{min}}^{z_{max}} \mathrm{d}z I_2[y,z] g(z)\,
 = \frac{2}{(1+r)^{2-2\epsilon}}\,
 y^{-\epsilon} \int_{z_{min}}^{z_{max}} \mathrm{d}z \,
\frac{ y_{max}^{\epsilon}(y_{max}-y)^{-\epsilon}}{(1-z)^{2+\epsilon}}\,
g(z) \,.
\eea
We note that here $g(z)$ can be replaced by $g(1)$ since those 
terms behave like $\int \mathrm{d}1/(1-z) \propto \log(y)$ are 
non-singular.

Then we use
\bea
&&\int_0^1\mathrm{d}y \int_{z_{min}}^{z_{max}}\mathrm{d}z \,
I_2[y,z](f(y)-f(0)) \,
+ f(0)\int_0^1\mathrm{d}y \int_{z_{min}}^{z_{max}}\mathrm{d}z I_2[y,z] \, \nn \\
&& \hspace{6.ex} = \frac{2}{(1+r)^2}\,
 \int_0^1 \mathrm{d}y \int_{z_{min}}^{z_{max}}\mathrm{d}z \,
\frac{1}{(1-z)^2} (f(y)-f(0))  \,
 + f(0) \int_0^1\mathrm{d}y \int_{z_{min}}^{z_{max}}\mathrm{d}z I_2[y,z] \nn \\
&& \hspace{6.ex} = \int_0^1 \mathrm{d}y 
\left(2 \kappa \left( \frac{1}{\kappa y} \right)_+ \,
\delta(1-z) -\frac{1}{\epsilon}\delta(y)\delta(1-z) \right)f(y)\,.
\eea
Again, we have expand $z_{max}$ and $z_{min}$ around $y=0$ and throw away
regular contributions. 

Therefore 
\bea
&&  \int_{z_{min}}^{z_{max}} \mathrm{d}z I_2[y,z] g(z) \nn \\
&=& \int_0^1 \mathrm{d}z \,
\left(-\frac{1}{\epsilon}\delta(y)\delta(1-z) \,
+2\kappa \left(\frac{1}{\kappa y}\right)_+ \delta(1-z) \right) g(z) \,.
\eea

{}


\begin{thebibliography}{}


\bibitem{Beneke:2000xiv}
M.~Beneke I.~Efthymiopoulos, M.~L.~Mangano and et. al., arXiv:hep-ph/0003033v1

\bibitem{Kharchilava:2000plb}
A.~Kharchilava, Phys. Lett. B {\bf 476} 73 (2000).



\bibitem{Bauer:2000prd}
C.~W.~Bauer, S.~Fleming and M.~Luke, Phys. Rev. D {\bf 63} 014006 (2000).

\bibitem{Bauer:2001prd}
C.~W.~Bauer, S.~Fleming,D.~Pirjol and I.~W.~Stewart, Phys. Rev. D {\bf 63} 114020 (2001).


\bibitem{Bauer:2001plb}
C.~W.~Bauer and I.~W.~Stewart, Phys. Lett. B {\bf 516} 134 (2001).

\bibitem{Bauer:2002prd}
C.~W.~Bauer, D.~Pirjol and I.~W.~Stewart, Phys. Rev. D {\bf 65} 054022 (2002).

\bibitem{HERWIG}
G. Corcella, I.G. Knowles, G. Marchesini, S. Moretti, K. Odagiri, P. Richardson, M.H. Seymour and B.R. Webber, JHEP {\bf 0101} 010 (2001) 


\bibitem{Procura:2010prd}
M.~Procura and I.~W.~Stewart, Phys. Rev. D {\bf 81} 074009 (2010).


\bibitem{Campbell:2004prd}
J.~Campbell, R.~K.~Ellis and F.~Tramontano, Phys. Rev. D {\bf 70} 094012 (2004).

\bibitem{Manohar:2007prd}
A.~V.~Manohar and I.~W.~Stewart, Phys. Rev. D {\bf 76} 074002 (2007).

\bibitem{Fleming:2006prd}
S.~Fleming, A.~K.~Leibovich and T.~Mehen, Phys. Rev. D {\bf 74} 114004 (2006).


\bibitem{Leibovich:2000prd}
A.~K.~Leibovich, I.~Low and I.~Z.~Rothstein, Phys. Rev. D {\bf 61} 053006 (2000).

\bibitem{Becher:2006prl}
T.~Becher and M.~Neubert, Phys. Rev. Lett. {\bf 97} 082001 (2006).

\bibitem{Fleming:2008prd}
S.~Fleming, A.~H.~Hoang, S.~Mantry and I.~W.~Stewart, Phys. Rev. D {\bf 77} 114003 (2008).


\bibitem{Leibovich:2001plb}
A.~K.~Leibovich, I.~Low and I.~Z.~Rothstein, Phys. Lett. B {\bf 513} 83 (2001). 




\end{thebibliography}
\end{document}